\documentclass[twocolumn]{svjour3}          
\smartqed  
\usepackage{graphicx}
\journalname{Journal of Superconductivity and Novel Magnetism}
\begin{document}

\title{Novel Spin-Orbital Phases Induced by Orbital Dilution
}

\author{Wojciech Brzezicki \and Mario Cuoco \and Andrzej M. Ole\'s
}

\authorrunning{W. Brzezicki, M. Cuoco and A.M. Ole\'s}

\institute{Wojciech Brzezicki
           \and
           Mario Cuoco \at
             CNR-SPIN, IT-84084 Fisciano (SA), Italy, and
             Dipartimento di Fisica \textquotedblleft{}E. R. Caianiello\textquotedblright{},
             Universit\'a di Salerno, IT-84084 Fisciano (SA), Italy
           \and
           Andrzej M. Ole\'s \at
             Marian Smoluchowski Inst. of Physics, Jagiellonian University,
             prof. S. \L{}ojasiewicza 11, PL-30348 Krak\'ow, Poland;
             Max Planck Institute for Solid State Research,
             Stuttgart, Germany
             \email{a.m.oles@fkf.mpg.de}
}

\date{Received: date / Accepted: date}

\maketitle

\begin{abstract}
We demonstrate that magnetic $3d$ impurities with $S=3/2$ spins and no
orbital degree of freedom induce changes of spin-orbital order in a
$4d^4$ Mott insulator with $S=1$ spins. Impurities act either as
spin defects which decouple from the surrounding ions, or trigger
orbital polarons along $3d$-$4d$ bonds. The $4d$-$4d$ superexchange
in the host $J_{\rm host}$ competes with $3d$-$4d$ superexchange
$J'_{\rm imp}$ --- it depends on which orbital is doubly occupied.
The spin-orbital order within the host is totally modified at
doping $x=1/4$. Our findings provide new perspective for future
theoretical and experimental studies of doped transition-metal oxides.
\end{abstract}

\section{ Orbital dilution }

Entangled spin-orbital superexchange interactions lead to several
surprises in transition-metal oxides \cite{Ole12}. In Mott insulators
these interactions are modified by doping which may generate novel
phases --- they emerge from the interplay of complex
spin-orbital-charge couplings. For instance, hole doping of a
ferromagnetic (FM) system with $d^1$ ionic configurations removes
locally orbital degrees of freedom and generates stripe phases with
orbital polarons \cite{Wro10}. We show below that similar polarons
emerge also in doped spin-orbital systems.

\begin{figure}[t!]
\begin{center}
\includegraphics[width=7.7cm]{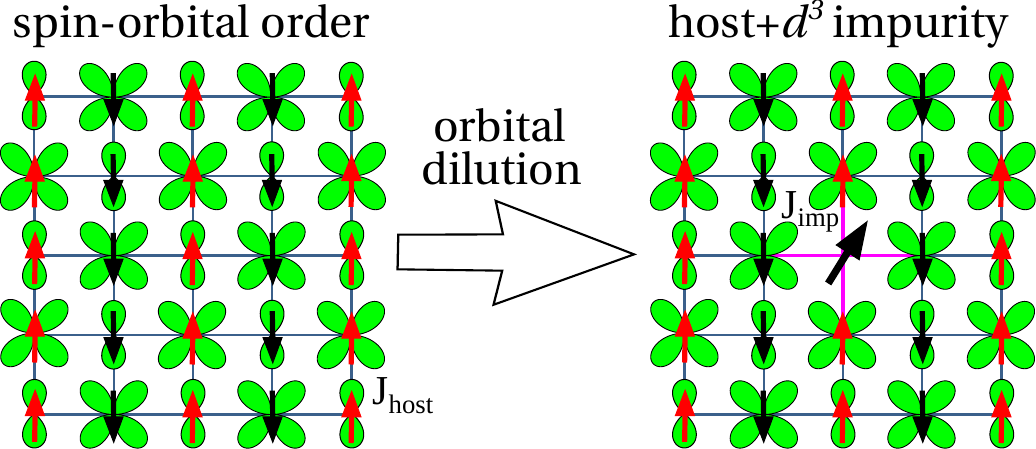}
\end{center}
\caption{
Left --- spin-orbital $C$-AF/$G$-AO order found in Mott insulators
with $d^2$ (vanadates) or $d4$ (ruthenates) ionic configurations. Spins
$S=1$ are shown by red (black) arrows and doubly occupied $t_{2g}$
orbitals (doublons) are shown by green symbols for $a$ and $c$
orbitals. Right --- the orbital dilution occurs after doping by a
$3d^3$ ion with $S=3/2$ and no orbital degree of freedom. Here $a$
doublon site is replaced by Mn$^{4+}$ or Cr$^{3+}$ impurity coupled
to the host by $3d$-$4d$ bonds.
\label{fig:bond}}
\end{figure}

The well known example are hole doped La$_{1-x}$Sr$_x$MnO$_3$
manganites with colossal magnetoresistance \cite{Dag01}. At low
doping orbital polarons emerge in an antiferromagnetic (AF) system by
double-exchange mechanism \cite{Kil99}, while at higher doping spin
order changes globally to a FM metal coexisting with an $e_g$ orbital
liquid phase \cite{Fei05}. But a different scenario is also possible
--- frustrated spin-orbital interactions may lead in some cases to the
collapse of any long range order and to a spin-orbital liquid suggested
for LiNiO$_2$ \cite{Ver04}.
However, spin and orbital energy scales are here quite different and
the reasons behind the absence of magnetic long range order are
indeed more subtle and not yet fully understood \cite{Rei05}.

A rather unique example of a spin-orbital system are perovskite
vanadates, where interesting competition between two types of
spin-orbital order was observed \cite{Fuj10}. The theoretical
spin-orbital model describes an interplay between $S=1$ spins and
$\tau=1/2$ orbital doublet $\{yz,zx\}$ active along the $c$ axis
\cite{Kha04}, while $xy$ orbitals are filled by one electron each.
In addition, $G$-type AF ($G$-AF) order in YVO$_3$ is fragile and
switches to $C$-type AF ($C$-AF) one in Y$_{1-x}$Ca$_x$VO$_3$
\cite{Fuj08}, with staggered lines of one-dimensional (1D) FM order
(Fig. 1), accompanied by $G$-type alternating orbital ($G$-AO) order.
Already at low doping $x\simeq 0.02$ finite spectral weight is
generated within the Mott-Hubbard gap due to charge defects
away from the VO$_6$ octahedra \cite{Ave13}.

Here we consider doping by static charge defects, e.g. Mn$^{4+}$ or
Cr$^{3+}$ ions, within the $ab$ planes having columnar $C$-AF order
of $S=1$ spins at ruthenium or vanadium ions in the host which leads to
\textit{orbital dilution} \cite{Brz15}, see Fig. 1. In contrast to hole
doped manganites, holes are here immobile and disturb $t_{2g}$ orbital
order. It has been shown that:
($i$) dilute Cr doping for Ru reduces the temperature of the
orthorhombic distortion and induces FM behavior in
Ca$_2$Ru$_{1-x}$Cr$_x$O$_4$ (with $0<x<0.13$) \cite{Qi10};
($ii$) Mn-substituted single crystals of Sr$_3$Ru$_{2-x}$Mn$_x$O$_7$
reveal an unusual $E$-type AF structure at $x=0.16$ \cite{Mes12}
which is again triggered by double exchange. These findings motivate
the theoretical search for the consequences of orbital dilution.

\section{ Isolated impurities }

For Ca$_2$RuO$_4$ host we use the spin-orbital model derived for
$t^4_{2g}$ ions with low ($S=1$) spins \cite{Cuo06}. This model uses
$t_{2g}$ doublons as orbital degrees of freedom and is isomorphic to
the vanadate $d^2$ model \cite{Kha04}, with the doublons transforming
into empty orbitals (occupied by two holes). We shall label $t_{2g}$
orbitals by index $\gamma$ when a given orbital is inactive along a
direction $\gamma\in\{a,b,c\}$:
\begin{equation}
\left|a\right\rangle\equiv\left|yz\right\rangle, \quad
\left|b\right\rangle\equiv\left|xz\right\rangle, \quad
\left|c\right\rangle\equiv\left|xy\right\rangle.
\label{eq:or_defs}
\end{equation}
We consider a two-dimensional (2D) square lattice with transition-metal
ions connected via oxygen orbitals as in an $ab$ RuO$_2$ plane of
Ca$_2$RuO$_4$ (SrRuO$_3$). In this case $|a\rangle$ ($|b\rangle$)
orbitals are active along the $b$ ($a$) axis, while $|c\rangle$
orbitals are active along both axes, $a$ and $b$. The superexchange in
the host for the bonds $\langle ij\rangle$ along the $\gamma\in\{a,b\}$
axis,
\begin{equation}
{\cal H}_{4d-4d}=J_{\rm host}\sum_{\langle ij\rangle\parallel\gamma}
\left\{J_{ij}^{(\gamma)}(\vec{S}_{i}\!\cdot\!\vec{S}_{j}+1)J_{ij}
+K_{ij}^{(\gamma)}\right\},
\label{eq:Hhost}
\end{equation}
is given by $J_{\rm host}$ and depends on orbital operators,
$J_{ij}^{(\gamma)}$ and $K_{ij}^{(\gamma)}$ \cite{Cuo06}. The above
form is generic \cite{Ole12} and the interactions depend on the
intraorbital Coulomb $U_2$ element and Hund's exchange $J_2$ in the
host.

We introduce two parameters to characterize the interactions along
the impurity-host $3d$-$4d$ bonds \cite{Brz15}:
\begin{equation}
J_{\rm imp}=\frac{t^{2}}{4\Delta}, \hskip 1.0cm
\label{eq:etai}
\eta_{\rm imp} =\frac{J_{1}}{\Delta}.
\label{eq:imp}
\end{equation}
Here the charge excitations $3d^3_i4d^4_j\Rightarrow 3d^4_i4d^3_j$
determine the $d^3$-$d^4$ superexchange \cite{Brz15} and involve the
energy
\begin{equation}
\Delta=I_{e}+3(U_{1}-U_{2})-4(J_{1}-J_{2}).
\label{Delta}
\end{equation}
It depends on the onsite Coulomb interactions $\{U_m\}$ ($m=1$
stands for the impurity $d^3$ ion),
on Hund's exchange $\{J_m\}$, and on the ionic energy $I_e$.
For Mn or Cr impurities in ruthenates $\Delta>0$.

\begin{figure}[t!]
\begin{center}
\includegraphics[width=7.5cm]{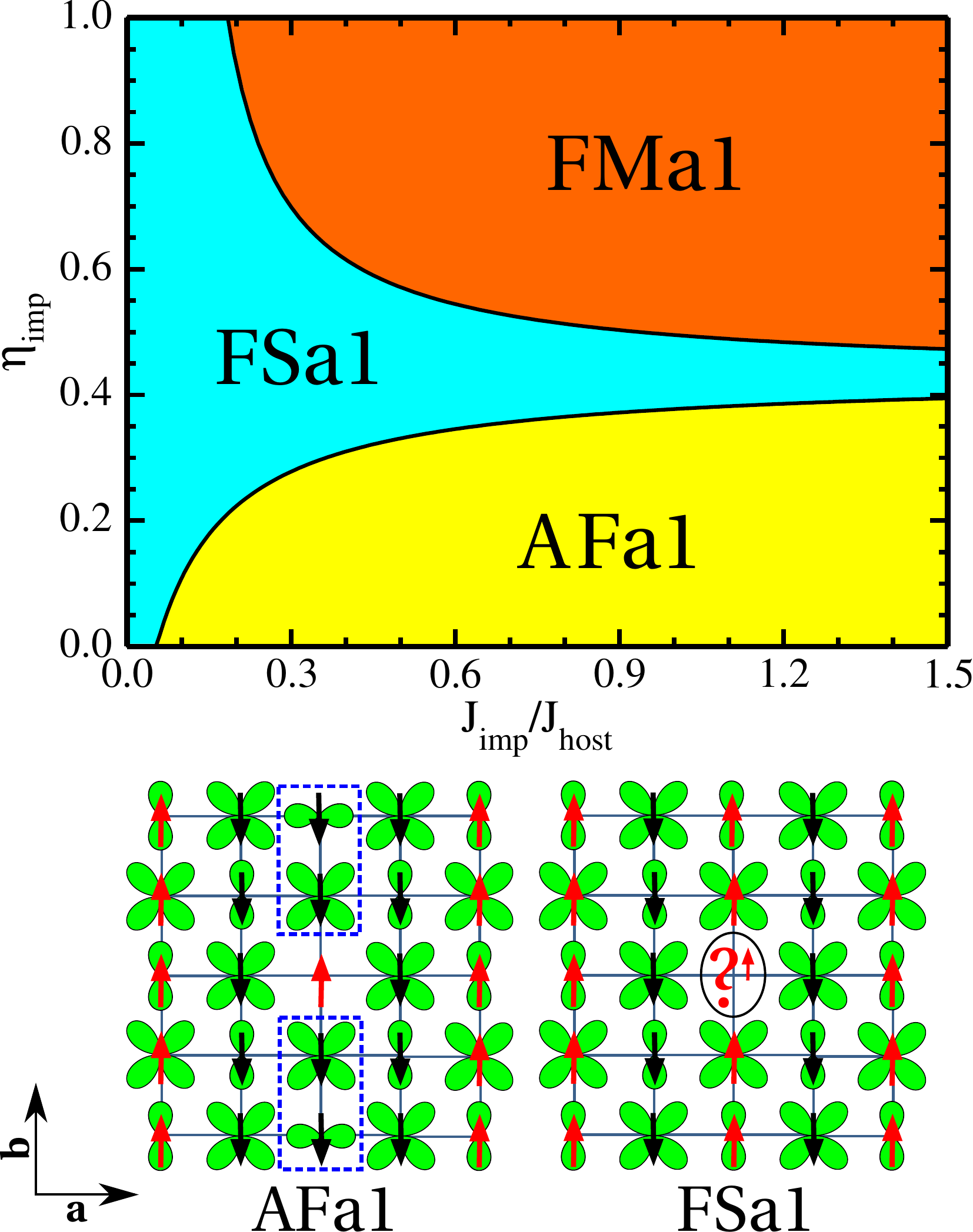}
\end{center}
\caption{
Phase diagram for the $3d$ impurity in the
$ab$ plane of $4d$ host with $C$-AF/$G$-AO order and impurity replacing
$a$ orbital doublon. Spin order is switched on the central vertical
line in phase AF$a1$ (dashed boxes), while in FS$a1$ phase the impurity
spin is frustrated and unable to alter $C$-AF pattern in the host; at
large $\eta_{\rm imp}$ FM order is found in FM$a1$ phase. }
\label{fig:one}
\end{figure}

With the parametrization introduced above, the dominant term in the
impurity-host Hamiltonian for the impurity spin $\vec{S}_i$ interacting
with the neighboring host spins $\{\vec{S}_j\}$ for nearest neighbors
$j\in{\cal N}(i)$ can be written in a rather compact form as follows
\begin{equation}
{\cal H}_{3d-4d}(i)\simeq\sum_{\gamma,j\in{\cal N}(i)}
\left\{J_{S}(D_j^{(\gamma)})(\vec{S}_i\!\cdot\!\vec{S}_j)
+ E_D D_j^{(\gamma)}\right\},
\label{eq:H123}
\end{equation}
where the spin couplings $J_S(D_j^{(\gamma)})$ depend on orbital
(doublon) configuration, $D_j^{(\gamma)}$ is the doublon projection
operator at site $j$ and the doublon energy $E_D$ depends on
$\eta_{\rm imp}$ Eq. (3). It can be shown \cite{Brz15} that the dominant
energy scale is $E_D^{\gamma}$, so for a single $3d$-$4d$ bond the
doublon does not occupy the inactive ($\gamma$) orbital and
spins couple with $J_{S}(D_j^{(\gamma)}=0)$ which can be either AF
if $\eta_{\rm imp}<0.43$ or FM if $\eta_{\rm imp}>0.43$. The change
of sign at $\eta_{\rm imp}^c\simeq 0.43$ is reminiscent of that found
in the Kugel-Khomskii model and as there \cite{Brz12} could lead to
exotic spin phases.

It has been found that a single $d^3$ impurity at site $i$ modifies the
spin-orbital order at its nearest neighbors $j\in{\cal N}(i)$, while
second nearest neighbors were assumed \cite{Brz15} to follow the
$C$-AF/$G$-AO order in the host (Fig. 1). Here we release this
constraint and consider classically the impurity with its first, second
and third nearest neighbors. One finds that the phase diagram is almost
unchanged by increasing cluster size from \cite{Brz15} for the doping
at $c$ doublon and doublon orbitals at nearest neighbor sites change
from inactive to active ones with increasing $J_{\rm imp}$ (not shown),
the latter similar to orbital polarons in manganites \cite{Dag04}.

In contrast, for $a$ doublon doping the impurity flips $a$ to $b$
orbitals at second neighbors along the vertical line and reverses
\textit{four} spins along the $b$ axis (replacing a 1D FM order along
if) in phase AF$a1$ for small $\eta_{\rm imp}$, see Fig. 2. This shows
that the modification of spin-orbital order induced by $3d$-$4d$ bonds
may be long range. Surprisingly, the host $C$-AF/$G$-AO order recovers
for larger $\eta_{\rm imp}\simeq\eta_{\rm imp}^c$ in frustrated spin
FS$a1$ phase when the AF $3d$-$4d$ coupling weakens. Here the quantum
fluctuations support the impurity spin following $C$-AF spin order.
Finally, when the impurity-host superexchange becomes strong FM, the
entire impurity cluster polarizes ferromagnetically in FM$a1$ phase.

\section{ Phase diagram at $x=1/4$ doping}

As for a single impurity, the $3d$-$4d$ bonds strongly influence
spin-orbital order at finite doping except for a relatively narrow
window of Hund's exchange $\eta_{\rm imp}\simeq\eta_{\rm imp}^c$.
Indeed, already for intermediate doping, $x=1/9$, $1/8$ or $1/5$,
spin-orbital order may change globally \cite{Brz15}. Here we
investigate higher periodic doping $x=1/4$, where half of the
superexchange bonds are $3d$-$4d$ hybrid ones and we show that
they dominate and dictate the overall spin-orbital order.

\begin{figure}[t!]
\begin{center}
\includegraphics[width=7.7cm]{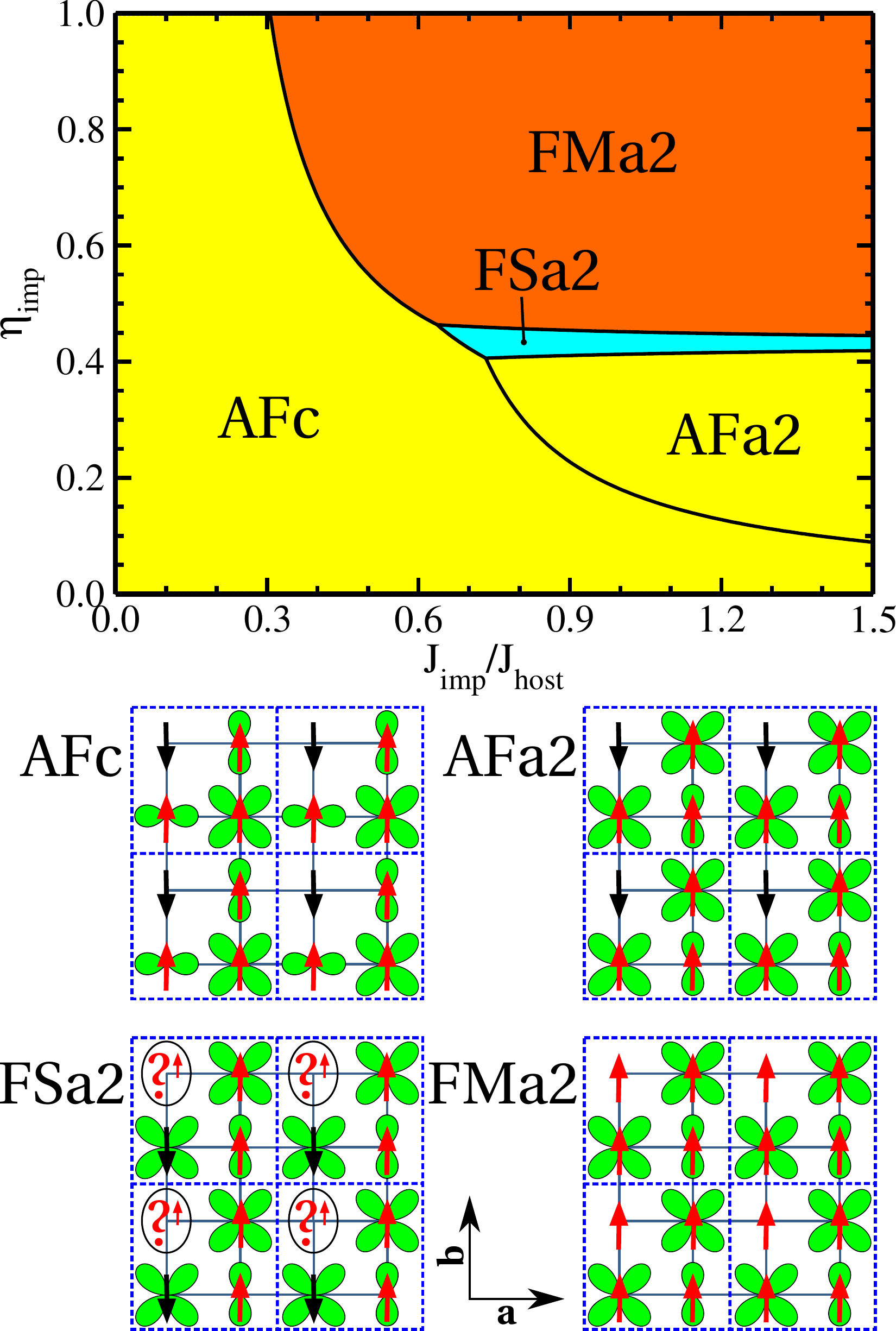}
\end{center}
\caption{
Ground state diagram obtained for periodic $x=1/4$ doping of $d^4$
spin-orbital host by $d^3$ ions (arrows). Dashed blue lines separate
$2\times 2$ unit cells in different phases; question marks in FS$a2$
indicate classically frustrated spins which order (small arrows) due
to quantum fluctuations.}
\label{fig:1to4}
\end{figure}

One finds that when at least one of the two impurity parameters is
small, either $J_{\rm imp}$ or $\eta_{\rm imp}$ (3), $c$ doublon sites
are doped and host-impurity coupling is AF. This AF coupling is
amplified by charge excitations on $3d$-$4d$ bonds when doublon
orbitals at $4d$ ions are inactive --- it dominates in this range of
the phase diagram and induces FM order along the $4d$-$4d$ host bonds
within the antiferrimagnetic (AFI) spin order for the entire $ab$ plane,
see Fig. 3. Thus only every second undoped vertical line is FM, as in
the initial $C$-AF phase, while host spins are inverted on any other
vertical line and the doublon orbitals flip from $a$ to $b$. Indeed,
this modification of the orbital order stabilizes the FM interactions
for $c$-$a$ doublon pairs on the horizontal bonds. For FM
order this orbital pattern is stabilized by double exchange,
similar to a hole doped $t_{2g}$ system \cite{Wro10}.
In contrast, if either $J_{\rm imp}/J_{\rm host}$ or $\eta_{\rm imp}$
is bigger than $\sim 0.5$, orbital dilution occurs on sublattice $a$.
When the $3d$-$4d$ superexchange is AF, the orbital order along the
undoped vertical FM lines is unchanged and one finds in phase AF$a2$
again the same spin order as in AF$c$, but now there are twice as
many $c$ doublons as $a$ ones.

As in the dilute limit (Fig. 2), impurity spins are frustrated also
at $x=1/4$ doping in phase FS$a2$ when
$\eta_{\rm imp}\simeq\eta_{\rm imp}^c$ (Fig. 3) and the $3d$-$4d$
superexchange nearly vanishes and changes sign. Then the host spins
in between experience almost entirely the AF $4d$-$4d$ superexchange
and follow $C$-AF order. Quantum fluctuations stabilize opposite to
them orientation of impurity spins, and this phase may be seen as a
precursor of the FM$a2$ phase which has again $75\%$ FM bonds as the
two AFI phases.
At sufficiently large $\eta_{\rm imp}$ FM order takes over in FM$a2$
phase (except for the range of $c$ doublon doping). The $G$-AO order
is the same in this latter phase as in the undoped host, see Fig. 3.

\section{ Discussion and summary }

Orbital dilution will play a role in several Mott insulators with
spin-orbital order doped by ions without active orbitals. We have
shown that the orbital order around impurities changes in general,
so even in the dilute limit one may expect observable effects due
to islands of reversed spins and doublon orbitals.

The phase diagram of Fig. 3 confirms the general rule  that only a
sufficiently strong coupling $J_{\rm imp}/J_{\rm host}$ leads to
orbital dilution on $a$ sublattice and to a rich competition between
various types of spin order \cite{Brz15}. We argue that the general
trends found here are generic and the phase diagrams are only
quantitatively modified by quantum fluctuations, at least in systems
with weak spin-orbit coupling. We have found that double exchange
leads to local or global changes of spin-orbital order, similar to
formation of orbital stripes \cite{Wro10} or orbital polarons in
doped manganites \cite{Kil99}. Such changes are expected to generate
novel spin-orbital-charge modulated patterns reported recently for
$t_{2g}$ systems \cite{Prl15}.

Summarizing, this study highlights the role played by orbital dilution
due to $d^3$ impurities in cubic spin-orbital systems and opens a new
avenue towards theoretical understanding of Mn-doped layered ruthenates
and related systems. We have shown that impurities change radically the
spin-orbital order around them in the entire parameter range.
As a general feature one finds frustrated impurity spins and their
ability to polarize the host orbitals around them. This property is
remarkable and concerns both the dilute and high doping regime, so one
expects global changes of spin-orbital order in doped materials.
It is challenging to investigate whether disordered impurities would
generate similar changes of spin-orbital order as well.

\textbf{Acknowledgments}
W.B. acknowledges support by the European Union's Horizon 2020
research and innovation programme under the Marie Sklodowska-Curie
grant agreement No. 655515.
M.C. acknowledges funding by the European Union FP7/2007-2013
programme, Grant Agreement No. 264098 -- MAMA.
We acknowledge support by Narodowe Centrum Nauki
(NCN, National Science Center), Project 2012/04/A/ST3/00331.

\end{document}